\documentstyle[prl,aps,twocolumn,epsfig]{revtex}
\begin{document}
\title{Measurement of the angular momentum of a rotating Bose-Einstein condensate}
\author{F. Chevy, K. W. Madison, and J. Dalibard}
\address{Laboratoire Kastler Brossel$^*$, D\'epartement de Physique de
l'Ecole Normale Sup\'erieure\\
24 rue Lhomond, 75005 Paris, France}
\date{\today}

\maketitle 

\begin{abstract}
We study the quadrupole oscillation of a Bose-Einstein condensate of $^{87}$Rb
atoms confined in an axisymmetric magnetic trap, after it has
been stirred by an auxiliary laser beam. 
The stirring may lead to the nucleation of one or more vortices,
whose presence is revealed unambiguously by the precession
of the axes of the quadrupolar mode. For a stirring frequency $\Omega$ below the single 
vortex nucleation threshold $\Omega_c$, 
no measurable precession occurs. Just above $\Omega_c$,
the angular momentum deduced from the precession is $\sim \hbar$. 
For stirring frequencies above $\Omega_c$ the angular momentum is a smooth and increasing function of $\Omega$,
until an angular frequency $\Omega_c'$ is reached at which the vortex lattice disappears and the
precession stops.
\end{abstract}
\vskip 5mm
\noindent Pacs: 03.75.Fi, 67.40.Db, 32.80.Lg
\vskip 5mm

The achievement of Bose-Einstein condensation in 
atomic gases has led to a new impulse in the study of quantum gases
\cite{Anderson95,Bradley957,Davis95,Fried98}.
Among the several questions which can be investigated in these systems
the properties of quantum vortices are some of the most intriguing and debated. 
A vortex is a singularity line 
around which the circulation of the velocity field is non zero. For a superfluid
this circulation is quantized and equal to $nh/M$, where $n$ is an integer and
$M$ the atomic mass \cite{Lifshitz}. 
These quantized vortices play an essential role in
macroscopic quantum phenomena, such as the superfluidity of liquid helium
\cite{Donnelly}
or the response of a superconductor to an external magnetic field
\cite{Tinkham}. In this paper
we determine the angular momentum of a gaseous Bose-Einstein 
condensate, as it is stirred near the vortex nucleation threshold. 
We show that the angular momentum $L_z$ per atom along the stirring axis
jumps from zero to  $\hbar$ (within experimental uncertainty), 
as expected from the quantization of the velocity field. We 
also measure the increase of $L_z$ as more vortices are nucleated in the system,
for a stirring frequency higher than the nucleation threshold.

We consider here a condensate harmonically trapped in a 
quasi-cylindrically 
symmetric potential. The angular momentum of the condensate
along the symmetry axis $z$ of the trap is 
measured using the frequencies of its collective excitations.
More precisely, as suggested in \cite{Zambelli98}, 
we study the two transverse quadrupole modes
carrying angular momentum along the $z$ axis of $m=\pm 2$
(see also \cite{Dodd97,Sinha97,Svidzinsky98}). 
In the absence of vortices, 
the frequencies $\omega_\pm/2\pi$ of these two modes are equal
as a consequence of the reflection symmetry about the $xy$ plane.
By contrast, for $L_z \neq 0$, this degeneracy is lifted by an amount:
\begin{equation}
{\omega_+ - \omega_-} = \frac{2\,L_z}{ M r_\perp^2} \ ,
\label{lift}
\end{equation} 
where $r_\perp^2$ stands for the average value of $x^2+y^2$ for the condensate.
Consequently the measurements of $\omega_+-\omega_-$ and of the
transverse size of the condensate provide the angular momentum
of the gas.

The result (\ref{lift}) is valid if 
the system is properly described by the hydrodynamic theory 
for superfluids \cite{Zambelli98}. This requires that the number of atoms $N$ is much larger
than the ratio $a_{\rm ho}/a$, where $a_{\rm ho}$ is the size of the ground state 
of the transverse motion in the trap and $a$ the scattering length decribing
the interactions between atoms at low temperature. For our experimental conditions
the quantity $Na/a_{\rm ho}$ is larger than 1000 which ensures the validity of
(\ref{lift}). The prediction (\ref{lift}) can be interpreted in
terms of the Sagnac effect: in presence of a vortex the rotation of the
condensate, in which the two excitations $m=\pm 2$ propagate, lifts their
degeneracy.

Two experiments have led so far to the observation
of a vortex line in a gaseous condensate \cite{Matthews99,Madison00}. 
The method used in  \cite{Matthews99} 
uses a combination of a laser and a microwave to print the desired 
velocity field onto the atomic wave function.
One  generates in this
way a condensate with atoms in a given internal state rotating around a second,
stationary condensate in another internal state.
The interference of these two condensates allows for a measurement of 
the $2 \pi$ phase shift of the condensate wave function around the vortex which 
 proves  the quantization of the circulation.
The second method \cite{Madison00}, which is used in the present paper, is a 
transposition of the rotating bucket experiment
performed on $\null^4{\rm He}$.
We superimpose onto the cylindrically symmetric 
magnetic potential a non-axisymmetric, dipole potential created
by a stirring laser beam. The combined potential leads to a cigar-shaped
harmonic trap with a slight anisotropic transverse profile.
The transverse anisotropy is rotated at an angular frequency 
$\Omega$ and can nucleate a vortex if above a 
critical frequency $\Omega_{\rm c}$.

The details of the experimental setup have been described in \cite{Madison00,Madison00b}.
For the preparation of the condensate we start with $10^9$ $^{87}$Rb atoms
in a Ioffe--Pritchard magnetic trap at a temperature  
$\sim 200\ \mu$K. The oscillation frequency of the atoms
along the longitudinal axis of the trap is 
$\omega_z/2\pi=10.3$~Hz.
For the results presented here, 
the transverse frequencies $\omega_\perp/2\pi$ have been varied between 170~Hz and 210~Hz 
by adjusting the bias field at the center of the trap\cite{anis}. 

An experimental sequence consists in four steps: (i) condensation {\it via} 
evaporative cooling, (ii) vortex nucleation,
(iii) excitation and evolution of the quadrupolar modes,
and (iv) characterization using absorption imaging after a time-of-flight expansion
of $T_{\rm tof}=25$~ms.  The probe laser for the imaging 
propagates along the $z-$axis, and the image gives the transverse 
$xy$ distribution of atomic positions after the expansion. 
From each image we extract 
the temperature of the cloud, the size of the condensate in the $xy$ plane, 
and the number of vortices which have been nucleated.  

We evaporatively cool the atoms with a radio-frequency sweep. The condensation
threshold is reached at a temperature 550~nK.
We continue the evaporative cooling to a temperature below 80~nK at which point
approximately $3\times 10^5$ atoms are left in the condensate. This number is evaluated 
from the size of the condensate after expansion, 
assuming an initial Thomas-Fermi distribution \cite{threebody}.

After the end of the cooling phase
we switch on the stirring laser beam, 
which is parallel with the long axis of the condensate. 
The central position of this beam is varied in time with 
respect to the $x=y=0$ axis by two acousto-optic modulators. 
This temporal variation and the stirring light intensity are 
chosen such that the stirring laser creates on the trapped atoms
a dipole potential which is well approximated
by $M\omega_\perp^2(\epsilon_X X^2 +\epsilon_Y Y^2)/2$ with $\epsilon_X=0.05$
and $\epsilon_Y=0.15$. The $X,Y$ basis rotates at constant angular frequency
$\Omega$ with respect to the fixed $x,y$ basis. The stirring phase
lasts 900~ms which is well beyond the typical vortex nucleation time  
found experimentally to be about 450~ms \cite{difference}. During this phase 
the evaporation frequency is 
raised to a relatively large value (magnetic 
well depth equal to 2~$\mu$K) in order not to perturb 
the nucleation process, and we observe a slight heating of the 
cloud with a final temperature of $130\;(\pm 50)$~nK.

At the end of the vortex nucleation phase 
we excite a quadrupolar oscillation
using the dipole potential created by the stirring laser now
on a fixed basis ($X,Y=x,y$) and with a 10-times larger intensity. This
potential acts on the atoms for a $0.3$~ms duration, which is short compared
to the transverse oscillation period. The potential created
can be decomposed into (i) a part 
proportional to $x^2-y^2$, which excites the transverse quadrupole motion
of the condensate, and (ii) a part proportional to $x^2+y^2$ which excites
the transverse $m=0$ breathing mode which is not 
relevant for the present study \cite{Stringari96}.

The transverse quadrupolar mode excited
is a linear superposition of the $m=\pm 2$ modes.
The lift of degeneracy between the frequencies of 
these two modes causes a precession of the eigenaxes of the quadrupole mode
at an angular frequency given by $\dot \theta=(\omega_+-\omega_-)/2|m|=
(\omega_+-\omega_-)/4$. 
Therefore the measurement of $\dot \theta$ together with the size of the
condensate gives access to $L_z$.

To determine $\dot \theta$ we let the atomic cloud oscillate
freely in the magnetic trap for an adjustable period $\tau$ (between $0$ and $8$~ms)
after the quadrupole excitation. We then perform the time-of-flight $+$
absorption imaging sequence,
and we analyze the images of the condensate using a Thomas-Fermi type
distribution \cite{fit} as a fit function,  
with an adjustable ellipticity and adjustable axes in the $xy$ plane.

\begin{figure}
\begin{center}
\epsfig{figure=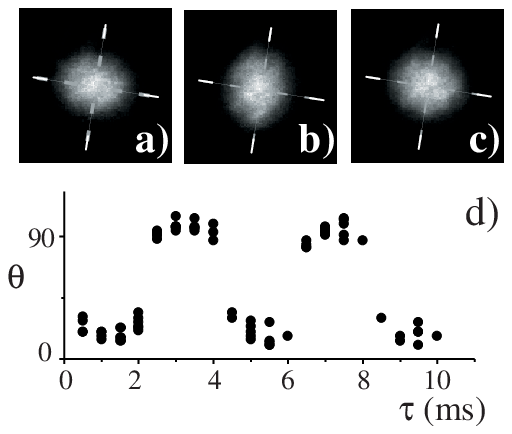,height=5.5cm}
\epsfig{figure=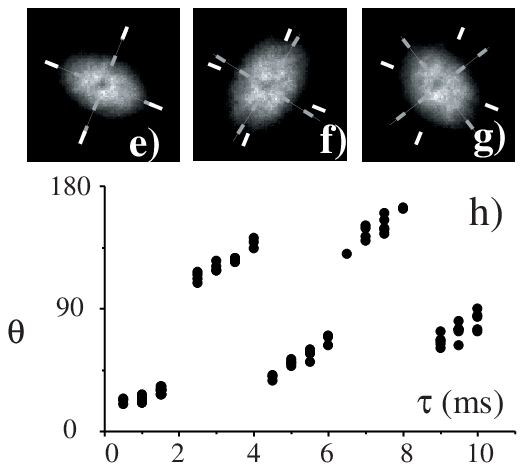,height=5.5cm}
\end{center}
\caption{Transverse oscillations of a stirred condensate 
with $N=3.7\;(\pm 1.1)\times 10^5$ atoms and $\omega_\perp/2\pi=171$~Hz. 
For a,b,c,d the sirring frequency is $\Omega/2\pi=114$~Hz,
below the vortex nucleation threshold $\Omega_c/2\pi=115$~Hz. For e,f,g,h, 
$\Omega/2\pi=120$~Hz. 
For a,e: $\tau=1$~ms; b,f: $\tau=3$~ms; c,g: $\tau=5$~ms. 
The fixed axes indicate the excitation basis and the rotating ones 
indicate the condensate axes.
A single vortex is visible at the center of the condensate in e,f,g.
The figures d and h give
the variations with $\tau$ of the direction $\theta$
of the large axis of the elliptical atomic cloud in the $xy$ plane.
Each circle represents a single realization of the experiment.
}
\label{fig1}
\end{figure}

A typical result is shown in fig. \ref{fig1}. For this measurement 
 the measured number of atoms was  $ 3.7\;(\pm 1.1)\times10^5$.
The transverse frequency  
$\omega_\perp/2\pi$ equals $171$~Hz,  and the threshold frequency 
$\Omega_c/2\pi$ for nucleating a vortex is $115$~Hz. The sequences of pictures fig. \ref{fig1}abc
and fig. \ref{fig1}def correspond to $\tau=1,3$ and 5~ms. Fig. \ref{fig1}abc 
has been taken after stirring the condensate at a frequency $\Omega/2\pi=114$~Hz.
Since $\Omega<\Omega_c$, no evidence for a vortex (which would appear as a 
density dip at the center of the condensate) is found in the corresponding images.
The quadrupole oscillation is identical to the one found in absence of the
vortex nucleation phase. As shown in fig. \ref{fig1}d the angle of the large axis of the 
elliptical condensate in the $xy$ plane oscillates between 0 and $\pi/2$ 
with the frequency $250$~Hz ($\pm 3$~Hz), which is in good agreement
with the value $\sqrt{2}\,( \omega_\perp/2\pi)$ expected for a zero  
temperature condensate in the Thomas-Fermi
limit \cite{Stringari96}. 

For a stirring frequency $\Omega/2\pi=120$~Hz for which
a vortex is systematically nucleated at the center of the condensate,
the behaviour of the system is dramatically different. A precession
of the long axis of the condensate occurs, as can be seen
in the sequence of pictures fig. \ref{fig1}efg. A measurement 
of the angle of the large axis of the condensate in the $xy$ plane as a 
function of $\tau$ is given in fig. \ref{fig1}h. It shows that
the axis of the condensate precesses, with an angular velocity of $\dot \theta$
equal to 5.9 ($\pm$~0.2) degrees per millisecond (
{\it i.e.} $(\omega_+-\omega_-)/2\pi=66$~Hz).

In order to deduce from this precession the value of $L_z$ we 
must determine the {\it in situ} size $r_\perp$ of the condensate. 
The fit of the image of the condensate after expansion 
gives an average radius equal to $103\;(\pm 6)~\mu$m 
(the radius is defined as the average distance from center
at which the Thomas-Fermi fit function vanishes). In our experimental conditions
the time-of-flight corresponds to a dilatation of the transverse lengths
by a factor $\sqrt{1 + \omega_\perp^2 T_{\rm tof}^2}=26.8$ 
\cite{Castin96,Kagan97}, so that the radius of the condensate
before time-of-flight is $R_\perp=3.8~\mu$m. The Thomas-Fermi
approximation yields $r_\perp^2=2\;R_\perp^2/7$ and we infer 
$L_z/\hbar=1.2\; (\pm 0.1)$ for fig. 1h. Quite remarkably a precise determination
of the number of atoms is not needed in this measurement of $L_z$.

We have performed similar experiments for several values of the stirring frequency
$\Omega$ and of the oscillation frequency $\omega_\perp$. 
For simplicity we have restricted our measurements to a single value of $\tau$, 
corresponding to the third maximum in the temporal evolution 
of the ellipticity of the cloud (e.g. $\tau=5.4$~ms for the parameters of fig. \ref{fig1}).
For each choice of $\Omega$ we measure 
$\theta$, from which we  deduce $\dot\theta \simeq \theta/\tau$, 
and the angular momentum.

\begin{figure}
\begin{center}
\epsfig{figure=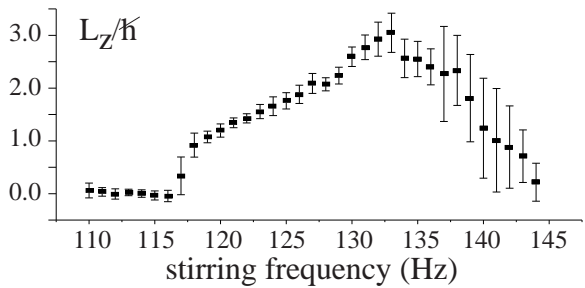,height=4.5cm}
\end{center}
\caption{Variation of the angular momentum deduced from (\ref{lift}) as a function
of the stirring frequency $\Omega$ for 
$\omega_\perp/2\pi=175$~Hz and $2.5\;(\pm 0.6)\times 10^5$ atoms.}
\label{fig2}
\end{figure}

\begin{figure}
\begin{center}
\epsfig{figure=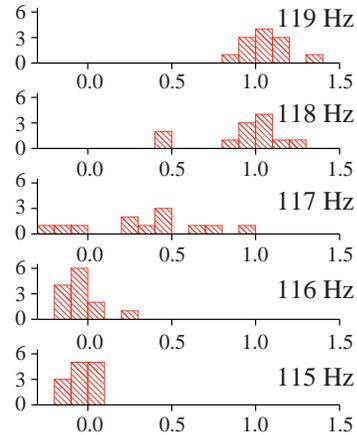,height=6.2cm}
\end{center}
\caption{Distribution of the values for the direction $L_z/\hbar$ of the large axis
of the condensate, after a 5~ms quadrupole oscillation for a stirring frequency
$\Omega/2\pi=115\ldots119$~Hz (same parameters as for fig. 2).}
\label{fig3}
\end{figure}

The results are shown in fig.~\ref{fig2} for the transverse frequency 
$\omega_\perp/2\pi=175$~Hz. These 
data allow for a  quantitative definition of the critical frequency
$\Omega_c$. This quantity was previously defined using a qualitative criterion:
the presence of a density dip at the center of the condensate. It can be 
defined now as the value for which $L_z$ jumps from 0 to a value close
to $\hbar$. The precision of this determination is of the order of
1~Hz, and it may be limited by the characteristic nucleation time
(450~ms). For $\Omega\sim\Omega_c$ the large error bar in 
fig.~\ref{fig2} reflects the large dispersion in the
results for $\theta$. This is illustrated more clearly in the histograms
of fig.~\ref{fig3} which give the value of $\theta$ for $\Omega/2\pi=115\ldots 119$~Hz.

For all transverse frequencies that we have achieved (between 170~Hz and 220~Hz)
we have found that $\Omega_c\simeq 0.65\;\tilde\omega_\perp$, where $\tilde\omega_\perp=\omega_\perp (1+(\epsilon_X+\epsilon_Y)/2)^{1/2}$ is the average transverse 
oscillation frequency in presence of the stirring laser. The  sensitivity of $\Omega_c$ 
to the atom number is very small: a change of $N$ by a factor of
$2$ changes $\Omega_c$ by less than 5\%. This is the reason for 
which a transition as sharp as that of fig. 2 is observed for 
$\Omega=\Omega_c$ although the relative dispersion of the atom 
number is 40\%. The measured value for $\Omega_c$ is notably 
larger (by $\sim 50\%$) than the predicted threshold above 
which the vortex state is energetically favored with respect 
to the non-vortex state \cite{Sinha97,Baym96,Lundh97,Fetter98,Castin99,Feder99}. 
A better account for the experimental result may be obtained by estimating 
the stirring frequency at which the energetic barrier 
between these two states disappears  
\cite{Dalfovo97,Isoshima99,Yvan00,Feder00}.

An important feature of fig.~\ref{fig2} is the linear variation
of the angular momentum as a function of $\Omega$ for $\Omega>\Omega_c$. 
We do not see a plateau for which $L_z$ stays 
constant and equal to $\hbar$
\cite{Butts99}. A possible interpretation may be that as soon
as the stirring frequency is large enough to nucleate a first vortex, 
a small change of $\Omega$ (by 1 
or 2 Hertz) is sufficient to nucleate more vortices. This is confirmed by 
a direct observation of the condensate images. For $\Omega/2\pi >124$~Hz, more
than 50\% of these images show clearly, around the center of the condensate, 
two or more density dips corresponding to 
vortices. Up to 5 vortices were observed on the images used
for fig.~\ref{fig2}.

Another remarkable feature appearing in fig.~\ref{fig2} concerns the
value of the angular momentum for stirring 
frequencies above a critical value $\Omega'_c$, 
when the images of the atomic cloud show a `turbulent' 
pattern with no evidence for a regular array of vortices
\cite{Madison00}. For the experimental conditions of fig. 2, 
$\Omega'_c/2\pi\sim 145$~Hz; at this stirring frequency, 
we find no precession of the eigenaxes of the quadrupolar oscillation,
corresponding to a zero angular momentum. The nature of
the atomic sample in this region of $\Omega$ is still an
open question: is it condensed or uncondensed? 
To address this question, we have performed for $\omega_\perp/2\pi=175$~Hz 
and $\Omega/2\pi=145$~Hz a series of measurements analogous to the ones
of fig.~\ref{fig1}dh and found that the quadrupole and monopole oscillation
frequencies were 256$\;(\pm 1)$~Hz and 350$\;(\pm 1)$~Hz 
respectively, which is  very close to what is expected 
of a zero temperature condensate at rest 
{\em i.e.} $\sqrt{2}\omega_\perp$ and $2\omega_\perp$.

To summarize we have presented in this paper 
a direct measurement of the angular momentum
of a Bose-Einstein condensate after it has been 
stirred at a given frequency $\Omega$. Using only 
``macroscopic" quantities such as the precession 
angle and the spatial extension of the condensate 
we have access to a ``microscopic" value of angular momentum $\sim\hbar$ per particule.
This measurement can be viewed 
as the transposition to gaseous condensates of the experiment performed 
with superfluid liquid helium by Vinen, which detected single quanta of circulation 
in rotating He II, by studying the lift of degeneracy between two vibrational modes
of a thin wire placed at the center of the rotating fluid \cite{Vinen}. 
Also the existence of the threshold $\Omega_c$ for the stirring frequency at which
the angular momentum per particle jumps from $0$ to $\hbar$ is a direct manifestation
of the superfluidity of the condensate, which complements 
the result of \cite{Raman99}, showing that an `object' moving
at a velocity below a critical value does not deposite any energy in a condensate.

{
We thank E. Akkermans, V. Bretin, Y. Castin, C. Cohen-Tannoudji, 
D. Gu\'ery-Odelin, C. Salomon,
G. Shlyapnikov, S. Stringari, and the ENS Laser cooling
group for several helpful discussions and comments. 
This work was partially supported by CNRS, Coll\`{e}ge de France,
DRET, DRED and EC (TMR network ERB FMRX-CT96-0002). This material is
based upon work supported by the North Atlantic Treaty Organization
under an NSF-NATO grant awarded to K.M. in 1999.

$^*$ Unit\'e de Recherche de l'Ecole normale sup\'erieure et de
l'Universit\'e Pierre et Marie Curie, associ\'ee au CNRS.
}
{\sl After this work was completed we became aware of an experiment 
similar to that of 
fig.1 by P. Haljan {\it et al}, QELS, May 7-12, postdeadline session.}


\begin{thebibliography}{}
\bibitem{anis} The measured residual asymmetry of the magnetic trap
is $\omega_{x,y}=\omega_\perp (1\pm 9.10^{-3})$.
\bibitem{difference} This procedure is slightly
different from the one in \cite{Madison00}, where the stirring
beam was switched on during the evaporation ramp, before the condensation
was reached. Both methods lead to similar results, 
in particular for $\Omega_c$. The method followed here gives
a slightly better control onto the number of atoms in the condensate.
\bibitem{fit} We use the usual ``inverted parabola" function \cite{Dalfovo99}
which is found to provide a good fit for the condensate image, except in the 
vicinity of the vortex core.
\end{thebibliography}
\end{document}